\documentclass[twocolumn,showpacs,amsmath,amssymb,prl]{revtex4}
\usepackage{amsxtra}
\usepackage{amssymb}
\usepackage{amsmath}
\usepackage{graphicx}
\usepackage[bf]{subfigure}

\begin{document}
\title{Phase-Conjugate Optical Coherence Tomography}
\date{\today}
\author{Baris I. Erkmen}
\email{erkmen@mit.edu}
\author{Jeffrey H. Shapiro}
\affiliation{Massachusetts Institute of Technology, Research Laboratory of Electronics, Cambridge, Massachusetts 02139, USA}

\begin{abstract}
Quantum optical coherence tomography (Q-OCT) offers a factor-of-two improvement in axial resolution and the advantage of even-order dispersion cancellation when it is compared to conventional OCT (C-OCT).  These features have been ascribed to the non-classical nature of the biphoton state employed in the former, as opposed to the classical state used in the latter.  Phase-conjugate OCT (PC-OCT), introduced here, shows that non-classical light is not necessary to reap Q-OCT's advantages.  PC-OCT uses classical-state signal and reference beams, which have a phase-sensitive cross-correlation, together with phase conjugation to achieve the axial resolution and even-order dispersion cancellation of Q-OCT with a signal-to-noise ratio that can be comparable to that of C-OCT.  
\end{abstract}
\pacs{07.60.Ly, 42.65.Lm, 42.50.Dv}
\maketitle

Optical coherence tomography (OCT) produces 3-D imagery through focused-beam scanning (for transverse resolution) and interference measurements (for axial resolution).  Conventional OCT (C-OCT) uses classical-state signal and reference beams, with a phase-insensitive cross-correlation, and measures their second-order interference in a Michelson interferometer \cite{Schmitt}.  Quantum OCT (Q-OCT) employs signal and reference beams in an entangled biphoton state, and measures their  fourth-order interference in a Hong-Ou-Mandel (HOM) interferometer \cite{Abouraddy,Abouraddy:two}.  In comparison to C-OCT, Q-OCT offers the advantages of a two-fold improvement in axial resolution and even-order dispersion cancellation.  Q-OCT's advantages have been ascribed to the non-classical nature of the entangled biphoton state, but we will report an OCT configuration that reaps both of these advantages with classical light.  

Q-OCT derives its signal and reference beams from spontaneous parametric down-conversion (SPDC), whose outputs are in a zero-mean Gaussian state, with a non-classical phase-sensitive cross-correlation function \cite{Sun, Shapiro:Gaussian}.  In the low-flux limit, this non-classical Gaussian state  becomes a stream of individually detectable biphotons.  Classical-state light beams can also have phase-sensitive cross-correlations, but quantum or classical phase-sensitive cross-correlations do not yield second-order interference.  This is why fourth-order interference is used in Q-OCT.  Our new OCT configuration---phase-conjugate OCT (PC-OCT)---uses phase conjugation to convert a phase-sensitive cross-correlation into a phase-insensitive cross-correlation that can be seen in second-order interference.  As we shall see, it is phase-sensitive cross-correlation, rather than non-classical behavior \em per se\/\rm, that provides the axial resolution improvement and even-order dispersion cancellation.

The basic block diagram for continuous-wave PC-OCT is shown in Fig.~1,  where we have suppressed all spatial coordinates, to focus our attention on the axial behavior, and we have drawn a transmission geometry, whereas the actual system would employ a bistatic geometry in reflection.  The signal and reference beams at the PC-OCT input are classical fields with a common center frequency $\omega_0$, and baseband complex envelopes, $E_S(t)$ and $E_R(t)$, with powers $\hbar\omega_0|E_K(t)|^2$, for $K = S,R$.  These complex fields are zero-mean, stationary, jointly Gaussian random processes that are completely characterized by their phase-insensitive auto-correlations
$\langle E_K^*(t+\tau)E_K(t)\rangle = {\cal{F}}^{-1}[S(\Omega)]$, for $K=S,R$, and their phase-sensitive cross-correlation $\langle E_S(t+\tau)E_R(t)\rangle = {\cal{F}}^{-1}[S(\Omega)]$, where 
\begin{equation}
{\cal{F}}^{-1}[S(\Omega)] \equiv \int_{-\infty}^\infty\!\frac{d\Omega}{2\pi}\,S(\Omega)
e^{-i\Omega\tau},
\end{equation}
is the inverse Fourier transform of $S(\Omega)$, and $S(\Omega) = S(-\Omega) \ge 0$ is the common spectrum of the signal and reference beams at detunings $\pm\Omega$ from $\omega_0$.  These fields have the maximum phase-sensitive cross-correlation that is consistent with classical physics \cite{Sun}.  

The signal beam is focused on a transverse spot on the sample yielding a reflection with complex envelope $E_{H}(t)= E_{S}(t)\star h(t)$, where $\star$ denotes convolution and  $h(t) = {\cal{F}}^{-1}[H(\Omega)]$
with
\begin{equation}
H(\Omega) = \int_0^\infty\!dz\, r(z,\Omega)e^{i 2 \phi(z, \Omega)}
\label{H:sample}
\end{equation}
being the sample's baseband impulse response.  
In Eq.~(\ref{H:sample}), $r(z,\Omega)$ is the complex reflection coefficient at depth $z$ and detuning $\Omega$, and $\phi(z,\Omega)$ is the phase acquired through propagation to depth $z$ in the sample.  
After conjugate amplification, we obtain the complex envelope
$E_{C}(t) = [E_{H}^{*}(t) + w(t)] \star \nu(t)$,
where $w(t)$, a zero-mean, circulo-complex, white Gaussian noise with correlation function $\langle w^*(t+\tau)w(t)\rangle = \delta(\tau)$, is the quantum noise injected by the conjugation process, and $\nu(t) = {\cal{F}}^{-1}[V(\Omega)]$ gives the conjugator's baseband impulse response in terms of its frequency response.  The output of the conjugator is refocused onto the sample resulting in the positive-frequency field
$E_{1}(t)= [E_{C}(t) \star h(t)]e^{-i\omega_0t}$, 
which is interfered with $E_2(t)  = E_R(t-T)e^{-i\omega_0(t-T)}$ in a Michelson interferometer, as shown in Fig.~1.  
\begin{figure}
\centering
\includegraphics[width=3in]{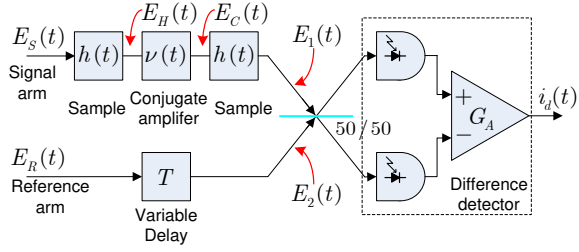}
\caption{(Color online) Phase-conjugate OCT.}
\end{figure}

The detectors in Fig.~1 are assumed to have quantum efficiency $\eta$, no dark current, and thermal noise with a white current spectral density $S_{i_{\rm th}}$.  The average amplified difference current, which constitutes the PC-OCT signature,  is then
\begin{eqnarray}
\langle i_d(t)\rangle &=& 2 q \eta G_A\,{\rm Re}\! \left(\int_{-\infty}^{\infty}\!  \frac{d\Omega}{2\pi}\,H^*\!(-\Omega) H(\Omega)\right. \nonumber \\[.12in]
&\times&\left.  V^*\!(-\Omega) S(\Omega) e^{-i(\Omega-\omega_0)T}   \right),
\label{TDOCT:signal}
\end{eqnarray}
where $q$ is the electron charge.

In C-OCT the signal and reference inputs have complex envelopes that are zero-mean, stationary, jointly Gaussian random processes which are completely characterized by their phase-insensitive auto- and cross-correlations,
$\langle E_J^*(t+\tau)E_K(t)\rangle = {\cal{F}}^{-1}[S(\Omega)]$, for $J,K=S,R$.
As shown in Fig.~2, C-OCT illuminates the sample with the signal beam and interferes the reflected signal---still given by convolution of $E_S(t)$ with $h(t)$---with the delayed reference beam in a Michelson interferometer.  Here we find that the average amplified difference current is 
\begin{eqnarray}
\lefteqn{\langle i_d(t)\rangle = 2 q \eta G_A} \nonumber \\[.12in]
&\times& {\rm Re}\! \left(\int_{-\infty}^{\infty}\!  \frac{d\Omega}{2\pi}\,H^*\!(-\Omega) S(\Omega)
e^{-i(\Omega-\omega_0)T}   \right).
\label{COCT:signal}
\end{eqnarray}
\begin{figure}
\centering
\includegraphics[width=2.5in]{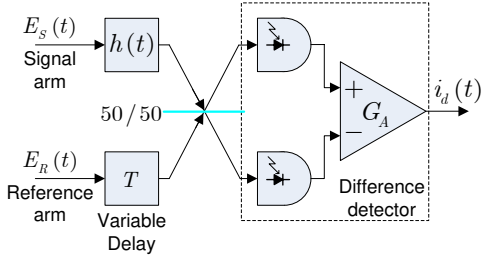}
\caption{(Color online) Conventional OCT.}
\end{figure}

For Q-OCT we must use quantum fields, because non-classical light is involved.  Now the baseband signal and reference beams are photon-units field operators, $\hat{E}_S(t)$ and $\hat{E}_R(t)$, with the following non-zero commutators, $[\hat{E}_J(t),\hat{E}_K^\dagger(u)] = \delta_{JK}\delta(t-u)$, for $J,K = S,R$.  Q-OCT illuminates the sample with $\hat{E}_S(t)$ and then applies the field operator for the reflected beam plus that for the reference beam to an HOM interferometer, as shown in Fig.~3.  The familiar biphoton HOM dip can be obtained theoretically---in a manner that is the natural quantum generalization of the classical Gaussian-state analysis we have used so far in this paper \cite{Sun}---by taking the signal and reference beams to be in a zero-mean joint Gaussian state that is completely characterized by the phase-insensitive (normally-ordered) auto-correlations
$\langle \hat{E}_K^\dagger(t+\tau)\hat{E}_K(t)\rangle = {\cal{F}}^{-1}[S(\Omega)]$,
for $K=S,R$, and the phase-sensitive cross-correlation
$\langle \hat{E}_S(t+\tau)\hat{E}_R(t)\rangle = {\cal{F}}^{-1}[\sqrt{S(\Omega)(S(\Omega) + 1)}]$.
This joint signal-reference state has the maximum possible phase-sensitive cross-correlation permitted by quantum mechanics.  In the usual biphoton limit wherein HOM interferometry is performed, $S(\Omega)\ll 1$ prevails, and the average photon-coincidence counting signature can be shown to be 
\begin{eqnarray}
\lefteqn{\langle C(T) \rangle = \frac{q^2 \eta^2}{2} \left [ \int_{-\infty}^{\infty}\!\frac{d\Omega}{2\pi}\, | H(\Omega) |^{2}  S(\Omega) \right.} \nonumber \\[.12in]
&-&  \left. {\rm Re}\! \left( \int_{-\infty}^{\infty}\!\frac{d\Omega}{2\pi} H^*\!(- \Omega) H(\Omega) S( \Omega)  e^{-i2\Omega T} \right) \right ]. 
\label{QOCT:signal}
\end{eqnarray}
\begin{figure}
\centering
\includegraphics[width=2.25in]{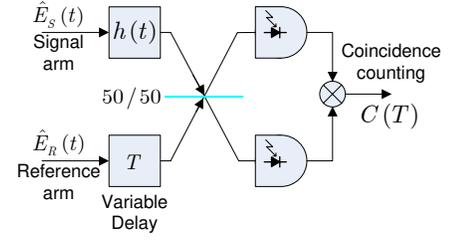}
\caption{(Color online) Quantum OCT.}
\end{figure}

Let us assume that $V^*(-\Omega)S(\Omega) \approx V^*S(\Omega) = (V^*P_S\sqrt{2\pi/\Omega_S^2})e^{-\Omega^2/2\Omega_S^2}$ and $H(\Omega) = re^{i(\omega_0+\Omega)T_0}$, with $|r|\ll 1$.  Physically, this corresponds to having a conjugate amplifier whose bandwidth is much broader than that of the signal-reference source, and a sample that is  a weakly-reflecting mirror at delay $T_0$.  Equation~(\ref{TDOCT:signal}) then gives a PC-OCT average amplified difference current that, as a function of the reference-arm delay $T$, is a sinusoidal fringe pattern of frequency $\omega_0$ with a Gaussian envelope proportional to $e^{-2\Omega_S^2(T_0-T/2)^2}$.  The average amplified difference current in C-OCT behaves similarly:  from Eq.~(\ref{COCT:signal}) we find that it too is a sinusoidal, frequency-$\omega_0$ fringe pattern in $T$, but its envelope is proportional to $e^{-\Omega_S^2(T_0-T)^2/2}$.    The signature of Q-OCT, found from Eq.~(\ref{QOCT:signal}), is a dip in the average coincidence-count versus reference-arm delay that is proportional to $e^{-2\Omega_S^2(T_0-T)^2}$.  Defining the axial resolutions of these OCT systems to be the full-width  between the $e^{-2}$ attenuation points in their Gaussian envelopes viewed as functions of $T_0$ shows that PC-OCT and Q-OCT both achieve factor-of-two improvements over C-OCT for the same source bandwidth.  

To probe the effect of dispersion on PC-OCT, C-OCT, and Q-OCT, we modify the sample's frequency response to $H(\Omega) = re^{i[(\omega_0+\Omega)T_0 + b\Omega^2/2]}$, where $b$ is a non-zero real constant representing second-order (group-velocity) dispersion.  Because the sample's frequency response enters the PC-OCT and Q-OCT signatures as $H^*(-\Omega)H(\Omega)$, neither one is affected by this dispersion term in $H(\Omega)$, i.e., it cancels out.  For C-OCT, however, we find that the Gaussian envelope of the average amplified difference current is now proportional to $e^{-\Omega_S^2(T_0-T)^2/2(1+\Omega^4_Sb^2)}$, i.e., its axial resolution becomes badly degraded when $\Omega_S^4b^2\gg 1$.  More generally, for $H(\Omega) = re^{i[(\omega_0+\Omega)T_0 + \beta(\Omega)]}$, PC-OCT and Q-OCT are immune to dispersion created by the even-order terms in the Taylor series expansion of $\beta(\Omega)$.  

Having shown that PC-OCT retains the key advantages of Q-OCT, let us turn to its SNR behavior.  Because Q-OCT relies on SPDC to generate the entangled biphoton state, and Geiger-mode avalanche photodiodes to perform photon-coincidence counting, its image acquisition is much slower than that of C-OCT, which can use bright sources and linear-mode detectors.  To assess the SNR of PC-OCT we shall continue to use the Gaussian spectrum for $S(\Omega)$ and the non-dispersing mirror for $H(\Omega)$, but, in order to limit its quantum noise, we take the conjugator's frequency response to be $V(\Omega) = Ve^{-\Omega^2/4\Omega_V^2}$.  We assume that $i_d(t)$ is time averaged for $T_I$\,sec (denoted $\langle i_d(t)\rangle_{T_I}$) at the reference-arm delay that maximizes the interference signature, and we define SNR$_{\text{PC-OCT}} = \langle i_d(t)\rangle^2/{\rm var}[\langle i_d(t)\rangle_{T_I}]$.   When the $w(t)$ contribution to the conjugator's output dominates the $E_H(t)$  contribution we find that
\begin{eqnarray}
\lefteqn{{\rm SNR}_{\rm PC-OCT} = } \nonumber \\[.12in]
&&\frac{8 T_I\eta |r|^4 |V|^2P_{S}^{2} \Omega_V^2/(\Omega_S^2 + 2\Omega_V^2)
 }    { \left [\Omega_{\rm th}+  P_{S} + |rV|^{2} \sqrt{\Omega_{V}^{2}/2\pi}    +\frac{\displaystyle 2 \eta |rV|^{2} P_{S} \Omega_{V}}{\sqrt{\displaystyle \Omega_{S}^{2}+\Omega_{V}^{2}}}\,  \right ]},
\label{SNR}
\end{eqnarray}
where $\Omega_{\rm th} \equiv S_{i_{\rm th}}/q^2 \eta$.  
From left to right the terms in the noise denominator are the thermal noise, the reference-arm shot noise,  the conjugate-amplifier quantum noise, and the intrinsic noise of the signal$\times$reference interference pattern itself.  Best performance is achieved when the conjugator gain $|V|^2$ is large enough to neglect the first two noise terms, and the input power $P_S$ is large enough that the intrinsic noise greatly exceeds the conjugator's quantum noise.  In this case we get
\begin{equation}
{\rm SNR}_{\text{PC-OCT}}  = 
\frac{4T_{I}|r|^2P_S\Omega_V\sqrt{\Omega_S^2 + \Omega_V^2}}{\Omega_S^2 + 2\Omega_V^2}.
\end{equation}

To compare the preceding SNR to that for C-OCT, we define ${\rm SNR}_{\text{C-OCT}} = \langle i_d(t)\rangle^2/{\rm var}[\langle i_d(t)\rangle_{T_I}]$ for the Fig.~2 configuration at the peak of the C-OCT interference signature.   When the reflected signal field is much weaker than the reference field, we then find that
\begin{equation}
{\rm SNR}_{\text{C-OCT}} = 4\eta T_I |r|^2P_S,
\end{equation} 
which can be \em smaller\/\rm\ than the ultimate ${\rm SNR}_{\text{PC-OCT}}$ result.  However, if PC-OCT's conjugator gain is too low to reach this ultimate performance, but its reference-arm shot noise dominates the other noise terms, we get
\begin{equation}
{\rm SNR}_{\text{PC-OCT}}  =  \frac{8\eta T_{I}|r|^4|V|^2P_S\Omega_V^2}{\Omega_S^2 + 2\Omega_V^2}, 
\end{equation}
which is substantially \em lower\/\rm\ than ${\rm SNR}_{\text{C-OCT}}$, because $|rV|^2 \ll 1$ is implicit in our assumption that the reference shot noise is dominant as high detector quantum efficiency can be expected.  Thus we can conclude that PC-OCT will have SNR similar to that of C-OCT, but only if high-gain phase conjugation is available \cite{SNRnote}.

At this juncture it is worth emphasizing the fundamental physical point revealed by the preceding analysis.  The use of entangled biphotons and fourth-order interference measurement in an HOM interferometer enable Q-OCT's two performance advantages over C-OCT:  a factor-of-two improvement in axial resolution and cancellation of even-order dispersion \cite{Abouraddy, Abouraddy:two}.  Classical phase-sensitive light also produces an HOM dip with even-order dispersion cancellation, but this dip is essentially unobservable because it rides on a much stronger background term \cite{Sun}.  Thus the non-classical character of the entangled biphoton is the source of Q-OCT's benefits, from which it might be concluded that non-classical light is required for any OCT configuration with these performance advantages over C-OCT.  Such is not the case, however, because our PC-OCT configuration shows that it is really phase-sensitive cross-correlations that are at the root of axial resolution enhancement and even-order dispersion cancellation.  Phase-sensitive cross-correlations cannot be seen in the second-order interference measurements used in C-OCT.  PC-OCT therefore phase conjugates one of the phase-sensitive cross-correlated beams, converting their phase-sensitive cross-correlation into a phase-insensitive cross-correlation that can be seen in second-order interference.  Our treatment of PC-OCT assumed classical-state light, and, because we need $S(0)\gg 1$ for high-SNR PC-OCT operation, little further can be expected in the way of performance improvement by using non-classical light in PC-OCT.  This can be seen by comparing the cross-spectra $S(\Omega)$ and $\sqrt{S(\Omega)(S(\Omega)+1)}$ when $S(\Omega) = (P_S\sqrt{2\pi/\Omega_S^2})e^{-\Omega^2/2\Omega_S^2}$ with $P_S\sqrt{2\pi/\Omega_S^2} \gg 1$.  

The intimate physical relation between PC-OCT and Q-OCT can be further elucidated by considering the way in which the sample's frequency response enters their measurement averages.  We again assume $V^*(-\Omega)S(\Omega) \approx V^*S(\Omega)$, so that both imagers yield signatures  $\propto \int\!d\Omega\, H^*(-\Omega)H(\Omega)S(\Omega)$.  Abouraddy \em et al\/\rm. \cite{Abouraddy} use Klyshko's advanced-wave interpretation \cite{Klyshko} to account for the $H^*(-\Omega)H(\Omega)$ factor in the Q-OCT signature as the product of an actual sample illumination and a virtual sample illumination. In our PC-OCT imager, this same $H^*(-\Omega)H(\Omega)$ factor comes from the two sample illuminations, one before phase conjugation and one after.  In both cases, it is the phase-sensitive cross-correlation that is responsible for this factor.  Q-OCT uses non-classical light and fourth-order interference while PC-OCT can use classical light and second-order interference to obtain the same sample information.  

That PC-OCT's two sample illuminations provide an axial resolution advantage over C-OCT leads naturally to considering whether C-OCT would also benefit from two sample illuminations.  Consider the Fig.~1 system with $E_S(t)$ and $E_R(t)$ arising from a C-OCT light source, and the phase-conjugate amplifier replaced with a conventional phase-insensitive amplifier of field gain $G(\Omega) = Ge^{-\Omega^2/4\Omega_G^2}$ with $|G|\gg 1$.   This two-pass C-OCT arrangement then yields an interference signature $\propto e^{-2\Omega_S^2(T_0-T/2)^2}$ for the weakly-reflecting mirror when the amplifier is sufficiently broadband, and an SNR given by Eq.~(\ref{SNR}) with $V$ replaced by $G$ and $\Omega_V$ replaced by $\Omega_G$.  Thus two-pass C-OCT has the same axial resolution advantage and SNR behavior as PC-OCT.  However, instead of providing even-order dispersion cancellation, two-pass C-OCT doubles all the even-order dispersion coefficients.     

Let us conclude by briefly addressing the implementation issues that arise with PC-OCT.  Our imager requires:  signal and reference light beams with a strong and broadband phase-sensitive cross-correlation; an illumination setup in which the signal beam is focused on and reflected from a sample, undergoes conjugate amplification, is refocused onto the same sample, and then interfered with the time-delayed reference beam; and a broadband, high-gain phase conjugator.  Strong signal and reference beams that have a phase-sensitive cross-correlation can be produced by splitting a single laser beam in two, and then imposing appropriate amplitude and phase noises on these beams through electro-optic modulators.  Existing optical telecommunication modulators, however, do not have sufficient bandwidth for high-resolution OCT.  A better approach to the PC-OCT source problem is to exploit nonlinear optics.  SPDC can have THz phase-matching bandwidths, and might be suitable for the PC-OCT application.  Unlike Q-OCT, which relies on SPDC for its entangled biphotons, a down-conversion source for PC-OCT can---and should---be driven at maximum pump strength, i.e., there is no need to limit its photon-pair generation rate so that these biphoton states are time-resolved by the $\sim$MHz bandwidth single-photon detectors that are used in Q-OCT's coincidence counter.  Hence pulsed pumping will surely be needed.  SPDC is also a possibility for the phase conjugation operation.  In a frequency-degenerate type-II phase matched down-converter, the reflected signal $E_H(t)$ is applied in one input polarization (call it the signal polarization) and a vacuum state field in the other (idler) polarization.  The idler output then has the characteristics needed for PC-OCT, viz., it consists of a phase-conjugated version of the signal input plus the minimum quantum noise needed to preserve free-field commutator brackets \cite{Shapiro:Gaussian}.  Similar phase-conjugate operation can also be obtained from frequency-degenerate four-wave mixing  \cite{Fisher, Feinberg, Norman}.  In both cases,  pulsed operation will be needed to achieve the gain-bandwidth product for high-performance PC-OCT.

In summary, we have a phase-conjugate OCT imager that combines many of the best features of conventional OCT and quantum OCT.  Like C-OCT, PC-OCT relies on second-order interference in a Michelson interferometer.  Thus it can use linear-mode avalanche photodiodes (APDs), rather than the lower bandwidth and less efficient Geiger-mode APDs employed in Q-OCT.  Like Q-OCT, PC-OCT enjoys a factor-of-two axial resolution advantage over C-OCT, and automatic cancellation of even-order dispersion terms.  The source of these advantages, for both Q-OCT and PC-OCT, is the phase-sensitive cross-correlation between the signal and reference beams.  In PC-OCT, however, this cross-correlation need not be beyond the limits of classical physics, as is required for Q-OCT.  Finally, PC-OCT may achieve an SNR comparable to that of C-OCT, thus realizing much faster image acquisition than is currently possible in Q-OCT.  All of these PC-OCT benefits are contingent on developing an appropriate source for producing signal and reference light beams with a strong and broadband phase-sensitive cross-correlation, and a phase conjugation system with suitably high gain-bandwidth product.   

\begin{acknowledgements}
J. H. Shapiro acknowledges useful technical discussions with F. N. C. Wong. This work was supported by the U. S. Army Research Office Multidisciplinary University Research Initiative Grant No.\@ W911NF-05-1-0197.
\end{acknowledgements}


\begin{thebibliography}{10}
\bibitem{SNRnote}Some care should be exercised in making this SNR comparison, because $\hbar\omega_0P_S$ is the total power that illuminates the sample in C-OCT, but it is only the initial sample illumination power in PC-OCT, i.e., there is also the power that illuminates the sample after the phase-conjugation operation.
\end{thebibliography}
\end{document}